# Host dependence of PL5 ensemble in 4H-SiC

Jiajun Li[1], Hui Qi[1], Feifei Zhou[1], Yumeng Song[1], Nanyang Xu[2], Bo Hong[1], Hongwei Chen[1, a)], Ying Dong[1, a)], Xinqing Wang[1, a)]

**AFFILIATIONS**

[1]College of metrology measurement and instrument, China Jiliang University, Hangzhou, 310018 China

[2]Institute of Quantum Sensing and College of Optical Science and Engineering, Zhejiang University, Hangzhou 310027, China

[a)]Authors to whom correspondence should be addressed: hwchen9814@cjlu.edu.cn, yingdong@cjlu.edu.cn; wxqnano@cjlu.edu.cn.

**ABSTRACT**

Color center PL5 in 4H silicon carbide (4H-SiC) has drawn significant attention due to its room-temperature quantum coherence properties and promising potential of quantum sensing applications. The preparation of PL5 ensemble is a critical prerequisite for practical applications. In this work, we investigated the formation of PL5 ensembles in types of 4H-SiC wafers, focusing on their suitability as hosts for PL5 ensemble. Results demonstrate that PL5 signals are exclusively observed in high-purity semi-insulating (HPSI) substrates, whereas divacancies PL1-PL4 can be detected in both HPSI and epitaxial samples. The type of in-plane stress in HPSI and epitaxial hosts is compressive in the same order of magnitude. Defects like stacking faults and dislocations are not observed simultaneously in the PL5 ensemble. Notably, the PL5 ensemble exhibits a relatively uniform distribution in the HPSI host, highlighting its readiness for integration into quantum sensing platforms. Furthermore, signal of PL5 can always be detected in the HPSI samples with different doses of electron irradiation, which suggests that HPSI wafers are more suitable hosts for the production of PL5 ensemble. This work provides critical insights into the material-specific requirements for PL5 ensemble formation and advances the development of 4H-SiC-based quantum technologies.

Divacancies in 4H silicon carbide (4H-SiC) have drawn great attentions in the field of quantum detections[1-3]. PL5 as one of the special divacancies, possesses room-temperature coherence properties, making it suitable to integrate with semiconductor devices for practical applications [4,5]. Production of PL5 ensemble is the basis for applications. Many effects have been made to produce PL5 samples. For example, single or ensemble of PL5 have been found in the divacancy arrays fabricated by focused ion beam or laser-writing both in epitaxial layers or HPSI wafers, though the yield of PL5 is much lower than that of PL1-PL4 [6-8]. PL5 ensembles can also be observed in the pristine HPSI samples or electron irradiated samples, whose photoluminescence intensities are relatively low and diminishes together with that of the regular divacancies PL1-PL4 with increasing measurement temperature in the same samples [9,10]. HPSI wafers are preferred to fabricate PL5 ensembles, whose reason is not clear yet.

PL5 is classified as one of special divacancies. However, the crystallographic structure of PL5 is unclear, making it hard for controllable production. One hypothesis of the crystallographic structure is that PL5 is a divacancy located in the stacking fault and stabilized by the quantum well, as suggested by the numerical calculations [11]. Yet another study suggests that PL5 is a divacancy slightly deviated from axial symmetry, based on the results of the electron paramagnetic resonance (EPR) on HPSI host wafers [12]. The conflicting conclusions of simulations and experiments hinder the production of PL5 ensembles.

In this work, types of 4H-SiC hosts have been compared to produce PL5 ensembles. The HPSI host is suitable to produce PL5 ensembles, while epitaxial layers or $n$-type wafers show no PL5 signal under liquid nitrogen environment. Processing factors have been discussed and shown no effect on the appearance of PL5. For HPSI host, the processing range for PL5 is narrower than that of divacancies PL1-PL4. Internal factors, including doping levels, inner stress and some defects commonly discussed for 4H-SiC, show no positive effects with the production of PL5 ensembles.

The host materials in this work were HPSI, epitaxial and $n$-type wafers, which were irradiated by the linear electron accelerator (10 MeV, $1\times10^{18}$ cm$^{-2}$, CIAE-DZ-10/20). Samples were then annealed in a vacuum tube furnace (<$5\times10^3$ Pa, OTF-1200X-S & GSL-1750-KS). The signal of divacancies was detected by photoluminescence (PL, 785 nm, 77 K, alpha 300R, WITec). Doping levels of samples were analyzed with secondary ion mass spectroscopy (CAMECA IMS 7f-Auto). The in-plane stress was calculated with the E2 transverse optical (TO) mode of hosts, detected by Raman spectroscopies (235 nm, Lab RAM HR Evolution, Horiba). The high-resolution transmission electron microscopy (TEM, Talos F200 & Themis Z) was used to observe the crystallographic structure of PL5 ensemble, whose sample was fabricated by the focused ion beam (FIB, Helios 5UX).

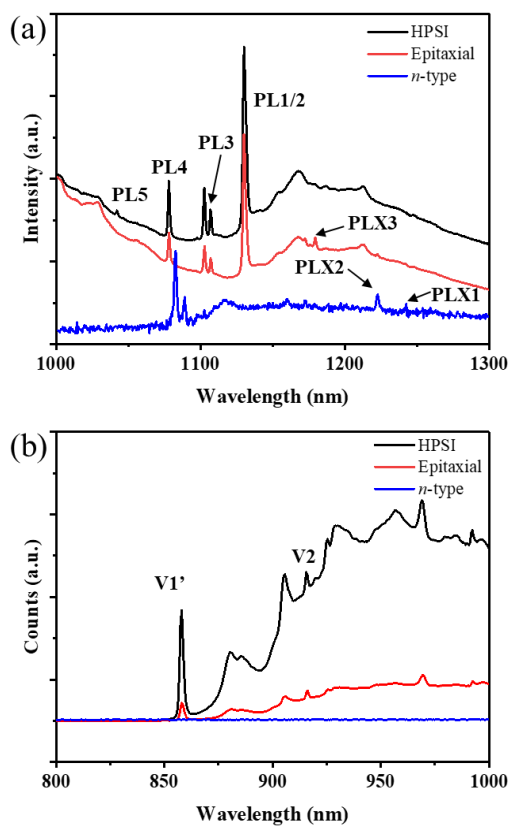

**FIG. 1.** PL spectra of (a) divacancies and (b) silicon vacancies in types of samples. The irradiation does is $1\times10^{18}$ cm$^{-2}$ and the annealing treatment is 1000 °C with 30 min.

To produce PL5, samples were annealed at 1000 °C for 30 min after $1\times10^{18}$ cm$^{-2}$ electron irradiation. Types of divacancies in different 4H-SiC hosts are detected at 77 K, as shown in Fig. 1(a). For the HPSI sample, divacancy peaks of PL1/PL2 (1132.0 nm/1130.5 nm), PL3(1130.5 nm), PL4(1078.5 nm), PL5 (1141.9 nm) can be detected. In the epitaxial sample, PL1-PL4 signals are observed. The signal of PL5 is absent, while one nitrogen vacancy (NV) peak PLX3 (1180.0 nm) arises. No divacancy can be detected in the *n*-type host but two NV center PLX1 (1242.8 nm) and PLX2(1223.1 nm).. As production of electron irradiation, two types of silicon vacancies in different hosts are detected, as shown in Fig. 1(b). Silicon vacancies V1' (858 nm) and V2 (916 nm) were detected in both HPSI and epitaxial hosts, while no signal in *n*-type host. Results show that silicon vacancies generated by electron irradiation have been transformed into other types of defects in *n*-type host. PL spectra agree well with former reports [10,13,14].

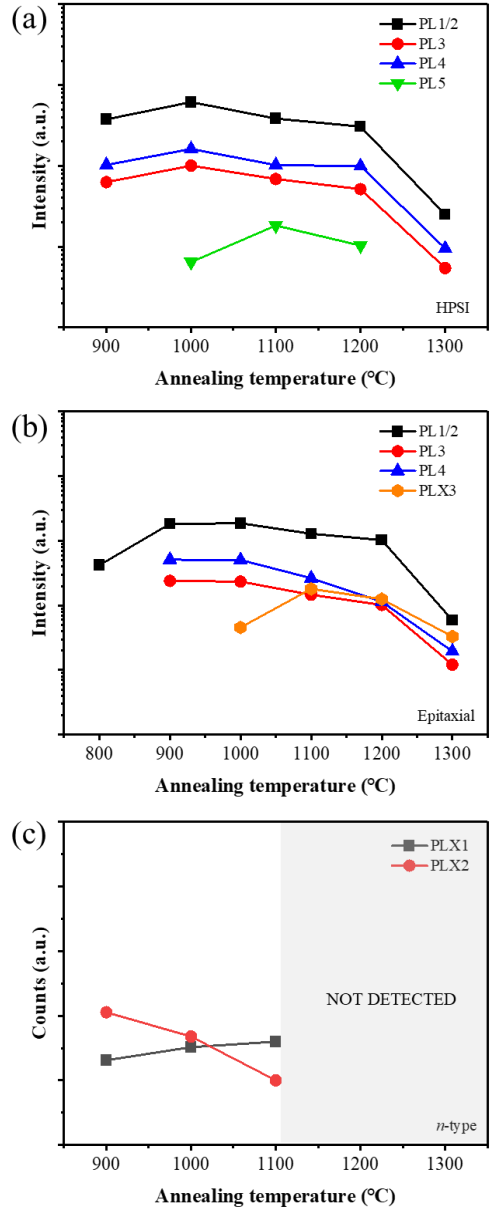

**FIG. 2.** Effect of annealing temperature on the divacancy intensities in (a) HPSI, (b) epitaxial and (c) *n*-type samples, the annealing temperature is 1000 °C and the duration is 30 min.

The annealing treatment was further investigated in samples with the annealing temperature ranging from 900 °C - 1300 °C. The annealing duration for all samples was 30 min. For HPSI samples, the signal of PL5 appears in the range of 1000 °C to 1200 °C and its intensity decreases at 1100 °C, as shown in Fig. 2(a). Meanwhile, PL1-PL4 signals are consistently detected in the range of 900 °C -1300 °C and their intensities decrease from 1000 °C. The formation process of divacancies is that the precursors migrate and combine with each other[15,16]. The annealing behavior of PL5 are in the same range of PL1-PL4, suggesting that there are similar dynamic properties between PL5 and PL1-PL4. However, the decrease temperature of PL5 intensity is higher than that of PL1-PL4, meaning that the driving force of PL5 formation is higher than that of PL1-PL4[17-19]. Meanwhile, the annealing temperature range of PL5 is narrower than that of PL1-PL4, the annihilation of PL5 is facile and the stability of PL5 is relatively low. Therefore, PL5 may possess some special structural properties compared with PL1-PL4. In the epitaxial samples, PL5 cannot be detected with the change of the annealing temperatures, although divacancy signals of PL1-PL4 are consistent, as shown in Fig. 2(b). Moreover, the nitrogen-vacancy center PLX3 comes along with PL1-PL4 starting from 1000 °C. Divacancy is not detected in *n*-type samples with various annealing processes, as shown in Fig. 2(c). The signals of nitrogen-vacancy centers PLX1 and PLX2 disappear from 1100 °C. Results show the HPSI and epitaxial wafers are suitable hosts for the PL1-PL4, while only HPSI samples are favored for the production of PL5 ensembles.

The common classification of wafers are the doping element and doping levels [20,21]. The concentrations of nitrogen in HPSI, epitaxial and *n*-type samples are $2\times10^{16}$ cm$^{-3}$, $1\times10^{16}$ cm$^{-3}$ and $1\times10^{18}$ cm$^{-3}$, respectively. The conductivity of *n*-type substrates are caused by heavily nitrogen doping[22,23]. As no divacancy has been detected, heavily doping of nitrogen is not beneficial for the formation of divacancies, including PL5. The epitaxial wafers are also conductive with much lower doping level than the *n*-type wafers, because of their low concentration of deep-level defects[24]. The doping level of the HPSI samples is in the same order of magnitude with the epitaxial samples. However, carriers in HPSI samples are trapped by the high concentration deep-level defects[25,26]. It has been demonstrated that the main deep-level defects in HPSI wafers are silicon/carbon vacancies or divacancies, which formed with considerable concentration during the high temperature growth process of physical vapor transportation [27-29]. However, silicon/carbon vacancies and divacancies can also form during the electron irradiation and annealing processes both in the HPSI and the epitaxial hosts [30-32]. The growth temperature of HPSI ingots is much higher than that of the epitaxial wafers, which means there are special defects with high formation energy generating in the HPSI hosts [33,34]. The special defects are connected with the formation of PL5, as PL5 ensembles formed only in the HPSI hosts.

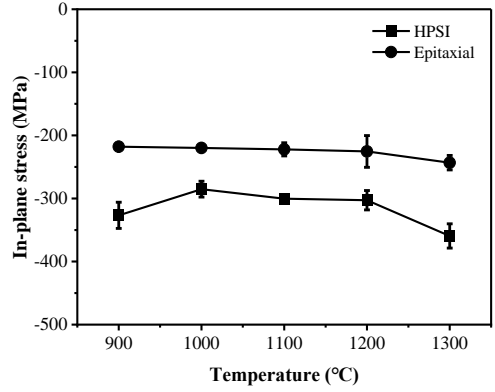

**Fig. 3.** In-plane stress calculated by Raman spectra E2(TO) with different annealing temperature. The annealing duration is 30 min for each sample.

The irradiation and annealing processes alter the stress in samples due to the lattice distortion during the treatments. For 4H-SiC, the E2 transverse optical (TO) mode in Raman spectroscopy corresponds to the Si-C vibration, which provides information about lattice distortion [35,36]. Therefore, the in-plane stress of crystals can be calculated by the shift of the E2(TO) peak [37]. The in-plane stress values of samples vary with annealing temperature ranging from 800 °C to 1300 °C. The stress values of samples increase initially and decrease from 1000 °C in HPSI samples, while it decreases monotonically in epitaxial samples, as shown in Fig. 3. The critical temperature for the stress transition in samples may be related to the dynamic variation of point defects [38]. The concentration of intrinsic point defects in HPSI samples is high, especially the deep-level defects, which can capture migrating defects to form multi-atoms defects, such as clusters [39-41]. There are high concentration of defects in the HPSI hosts [42]. With the increase of the annealing temperatures, point defects like silicon vacancies or interstitials induced by electron irradiation recombine at first, leading to the relief of stress initially. At higher annealing temperatures, more defects get activated and become mobile [43,44], accelerating the formation of multi-atom defect like clusters, which generates noticeable lattice distortion and elevate the stress values [45,46], leading to the increase of lattice distortion and the stress in crystals. However, the intrinsic concentration of point defects in epitaxial samples is much lower than that in HPSI samples [33]. It is easy to recombine for the point defects generated by the electron irradiation and the concentration of complex defects in epitaxial samples is much lower than that in HPSI samples. Therefore, the in-plane stress values of epitaxial samples decrease monotonically with increasing annealing temperature, as the distortion induced by electron irradiation is reduced. Meanwhile, the inner stress of epitaxial samples is always lower than that of HPSI samples. The type of stress is the same in HPSI and epitaxial samples, as well as the order of magnitude of the stress values. There is no direct correlation between the in-plane stress and the appearance of PL5.

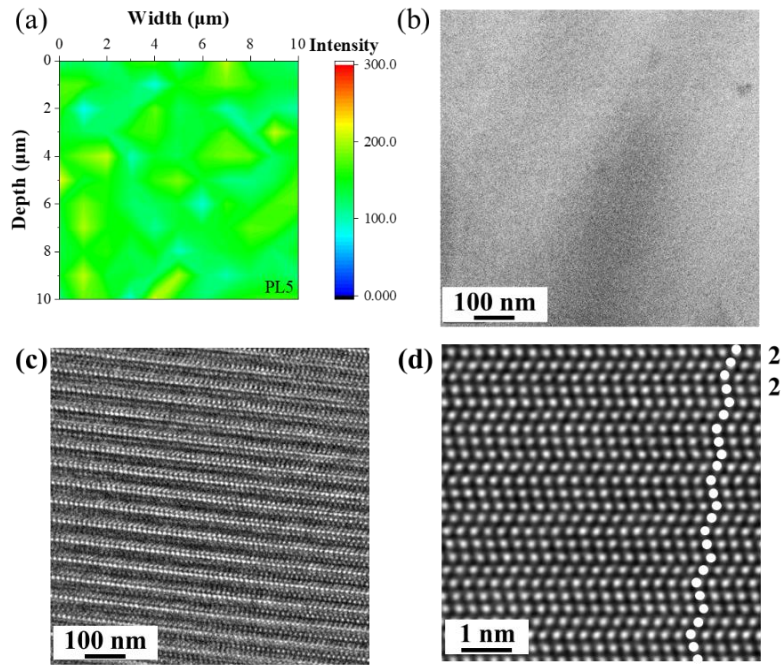

**Fig. 4.** (a) Mapping of PL5 signal intensity along *c*-axis. (b) PL spectra of different points at different depth. (c) Cross-sectional bright-field image of the mapping field of the marked area and (d) high-resolution TEM of the marked area.

With optimized annealing process, the PL5 ensemble was produced in the HPSI host. A 10 μm × 10 μm mapping with scanning step of 1 μm of PL5 signals along the *c*-axis of the sample is shown in Fig. 4(a). Although the intensity varies with different positions, PL5 signal can be detected at every pixel in the mapping image and the intensities at different positions are in the range of 100-200, meaning that PL5 distribution is uniform relatively in the host. 3 points at different depth are chosen to the spectra of divacancies, as shown in Fig.4(b), which verifies the distribution of PL5. The lattice microstructure of the mapped area was further characterized with TEM, which was prepared from the center of the mapped region. For PL5 ensemble, the stacking structure remains (2,2). There is no contrast of lattice distortion indicating stacking faults or dislocations observed in the cross-sectional bright-field image or in the high-resolution image, as shown in Fig. 4(c) and (d). In addition, the interplanar spacing of the (0001) plane in the sample was calculated from Fig. 4(d), yielding a value of 9.87 Å, which is smaller than the standard interplanar spacing of 10.82 Å for the (0001) plane in 4H-SiC [21].Therefore, it exhibits compressive stress in the host after irradiation and annealing, which is in consistent with the Raman analysis. Results of the microstructure analysis suggest that no observable defects are associated with the presence of PL5.

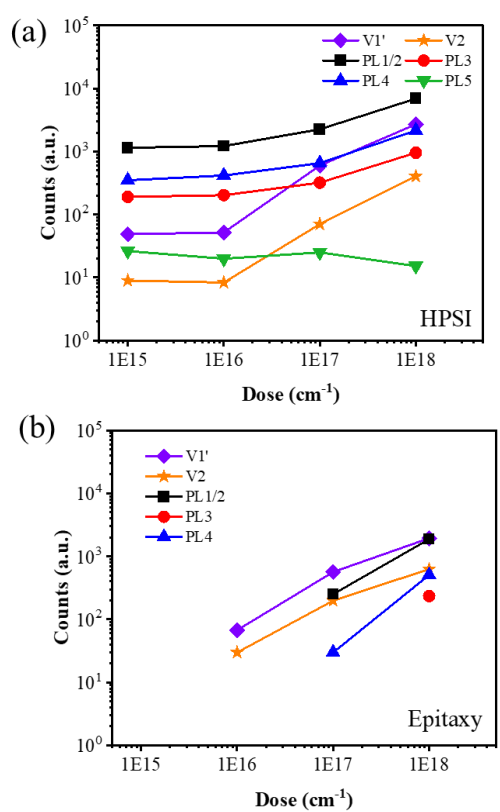

**Fig. 5.** (a) PL Comparation of the primary HPSI and epitaxial samples, and intensities of divacancies with different irradiation doses in (a) HPSI and (b) epitaxial samples.

The appearance of PL5 is detected exclusively in the HPSI sample in this work. To further investigate the host dependence of PL5 formation, samples with irradiation doses ranging from $1\times 10^{15}$ cm$^{-3}$ to $1\times 10^{18}$ cm$^{-3}$ were prepared. To verify the effect of irradiation, the intensities of divacancies PL1-PL4 and silicon vacancies V1' and V2 were also measured. In HPSI samples, divacancies PL1-PL5 and silicon vacancies V1' and V2 are consistently detected. The intensity of divacancies increases with increasing irradiation dose, as shown in Fig. 5(a). In epitaxial samples, V1' and V2 are detected starting at an irradiation dose of $1 \times 10^{16}$ cm$^{-3}$, while divacancies PL1/PL2/PL4 and PL3 are detected starting at an irradiation dose of $1\times10^{17}$ cm$^{-3}$ and $1\times10^{18}$ cm$^{-3}$, respectively. It indicates that HPSI host is easier to form divacancies, which corresponds with the analysis of annealing processes and stress variation. Therefore, PL5 is more readily formed in HPSI samples than in epitaxial samples. In addition, PL5 signals are consistently detected in HPSI samples but are absent in all epitaxial samples in this work. Given that the semi-insulating property of HPSI samples are attributed to their high concentration of intrinsic defects, some of these intrinsic defects are part of the influencing factors to form PL5.

In conclusion, the uniform PL5 ensemble has been produced with electron irradiation and annealing processes. The preferred host for PL5 ensemble is HPSI 4H-SiC wafers. Although both HPSI and epitaxial layers can be used as host materials to produce PL1-PL4. Multiple factors of treatment parameters and host features have been discussed to study the host preference for the fabrication of PL5. The types of in-plane stress show no effect on the production of PL5. The observable defects, such as stacking faults or dislocations, do not come along with the formation of PL5. It is deduced that intrinsic point defects of HPSI play important roles in the formation of PL5. This work proposes the method for the production of unform PL5 ensembles and paves the way for integrated quantum sensing based on 4H-SiC platform.

## ACKNOWLEDGEMENTS

This work is supported by the National Key Research and Development Program of China (Grant No. 2023YFF0718400) and the Major Program of Natural Science Foundation of Zhejiang province, China (Grant No. LD25A040001).

## AUTHORDECLARATIONS

### Conflict of Interest

The authors have no conflicts to disclose.

### Author Contributions

Jiajun Li and Hui Qi contributed equally to this work.

Jiajun Li: Conceptualization (equal); Data curation (equal); Formal analysis (equal); Investigation (equal); Methodology (equal); Validation (equal); Writing - original draft (equal). Hui Qi: Formal analysis (equal); Investigation (equal); Methodology (lead); Writing - original draft (supporting). Feifei Zhou: Data curation (equal); Investigation (equal); Writing original draft (supporting). Yumeng Song: Data curation (equal); Investigation (equal); Writing original draft (supporting). Bo Hong: Data curation (supporting); Investigation (supporting). Hongwei Chen: Data curation (supporting); Investigation (supporting); Supervision (equal); Writing– original draft (supporting). Ying Dong: Data curation (supporting); Investigation (supporting); Supervision (equal). Xinqing Wang: Conceptualization (equal); Funding acquisition (equal); Methodology (equal); Project administration (equal); Supervision (equal).

**DATA AVAILABILITY**

The data that support the findings of this study are available from the corresponding authors upon reasonable request.